\documentclass[journal]{IEEEtran}
\ifCLASSINFOpdf
\else
\fi
\usepackage{graphicx}  % <--- added
\usepackage{todonotes}
\usepackage[font=small,labelfont=bf]{caption}
\usepackage[left=0.5in, right=.3in, top=.5in, bottom=.3in]{geometry}
\begin{document}
\title{Real-time Detection and Adaptive Mitigation of Power-based Side-Channel Leakage in SoC}


\author{\IEEEauthorblockN{Pantea Kiaei\IEEEauthorrefmark{1},
Yuan Yao\IEEEauthorrefmark{2}, and
Patrick Schaumont\IEEEauthorrefmark{1}}\\
\IEEEauthorblockA{\IEEEauthorrefmark{1} Worcester Polytechnic Institute, Worcester, MA 01609 USA}\\
\IEEEauthorblockA{\IEEEauthorrefmark{2} Virginia Tech, Blacksburg, VA 24061 USA}% <-this % stops an unwanted space
\thanks{E-mails: pkiaei@wpi.edu; yuan9@vt.edu; pschaumont@wpi.edu}
}
% \author{\IEEEauthorblockN{Michael Shell\IEEEauthorrefmark{1},
% Homer Simpson\IEEEauthorrefmark{2},
% James Kirk\IEEEauthorrefmark{3}, 
% Montgomery Scott\IEEEauthorrefmark{3}, and
% Eldon Tyrell\IEEEauthorrefmark{4},~\IEEEmembership{Fellow,~IEEE}}
% \IEEEauthorblockA{\IEEEauthorrefmark{1}School of Electrical and Computer Engineering,
% Georgia Institute of Technology, Atlanta, GA 30332 USA}
% \IEEEauthorblockA{\IEEEauthorrefmark{2}Twentieth Century Fox, Springfield, USA}
% \IEEEauthorblockA{\IEEEauthorrefmark{3}Starfleet Academy, San Francisco, CA 96678 USA}
% \IEEEauthorblockA{\IEEEauthorrefmark{4}Tyrell Inc., 123 Replicant Street, Los Angeles, CA 90210 USA}% <-this % stops an unwanted space
% % \thanks{Manuscript received December 1, 2012; revised August 26, 2015. 
% % Corresponding author: M. Shell (email: http://www.michaelshell.org/contact.html).}
% }

\markboth{Boston (and Beyond) Area Architecture Workshop 2021}%
{}
% \markboth{Journal of \LaTeX\ Class Files,~Vol.~14, No.~8, August~2015}%
% {Shell \MakeLowercase{\textit{et al.}}: Bare Demo of IEEEtran.cls for IEEE Transactions on Magnetics Journals}



% for Transactions on Magnetics papers, we must declare the abstract and
% index terms PRIOR to the title within the \IEEEtitleabstractindextext
% IEEEtran command as these need to go into the title area created by
\IEEEtitleabstractindextext{%
\begin{abstract}
Power-based side-channel is a serious security threat to the System on Chip (SoC). The secret information is leaked from the power profile of the system while a cryptographic algorithm is running. The mitigation requires efforts from both the software level and hardware level. Currently, there is no comprehensive solution that can guarantee the whole complex system is free of leakage and can generically protect all cryptographic algorithms. In this paper, we propose a real-time leakage detection and mitigation system which enables the system to monitor the side-channel leakage effects of the hardware. Our proposed system has extensions that provide a real-time monitor of power consumption, detection of side-channel leakage, and real-time adaptive mitigation of detected side-channel leakage. Our proposed system is generic and can protect any algorithm running on it. 
\end{abstract}

}

\maketitle

\IEEEdisplaynontitleabstractindextext

\IEEEpeerreviewmaketitle
\vspace{-.3cm}
\section{Introduction}
% \section*{Motivation} 
%\noteBlue{Yuan}
%\noteRed{same as what we had for the presentation}
% backgroud about side-channel and it is a threat to the SoC
% \IEEEPARstart{S}{ide-channel} 
Side-channel leakage (SCL) discloses secret information through power consumption, electromagnetic dissipation, execution time, and other sources. In this work, our focus is on power-based SCL. Power side-channel analysis (SCA) has made the hardware attacker a particularly powerful adversary, relevant to embedded applications of secure System-on-Chips (SoC). When an SoC handles secret values, such as secret keys used in a symmetric cryptographic algorithm, the physical side-effects of these computations may be exploited as SCL. SCL is a critical vulnerability of secure SoCs as sophisticated SCA enables attackers to learn about the secret information even when the cryptographic algorithm is computationally secure.
% which can be applied both in-situ (CPA, DPA \cite{kocher1999differential, brier2004correlation}) and remotely \cite{zhao2018fpga},  . 
%\todo{should the explanation align with the two-ends explanation  in the abstract?}

When a program is running on a processor, unexpected SCL can happen. A software program is optimized and transformed into an executable file by a compiler. Generally, compilers are designed to optimize the speed or memory footprint of code, however, that is not always aligned with critical software security concerns. For example, in a masked implementation, compiler tasks such as register allocation and strength reduction (leading to shift operations), might cause unintended unmasking. Furthermore, even the existing algorithms for secure software design, like probing secure multiplication (ISW) \cite{ishai2003private} or bounded-moment secure multiplication \cite{barthe2017parallel}, have been shown to have leakage caused by unintended hardware effects  \cite{balasch2014cost, gregoire2018vectorizing}.
% because of operations that handle secret values. 
% The operations may include memory accesses, conditional branches, and arithmetic computations. 
While the software implementation causes the side-channel leakage, it is the processor hardware that creates the physical effects of SCL. Therefore, even if the software includes countermeasures against SCL, it is very hard to guarantee that the underlying hardware will be leakage free while running the software program. 
A processor consists of many elements including memory units and pipeline stages. Power SCL can arise from several parts of the micro-architecture simultaneously at times and individually at others.
% Pipelining is a common design method for increasing the processor's throughput and clock rate. However, pipelining leads to overlapped execution of instructions which makes it very difficult to pinpoint the source of leakage in a program. Security engineers can figure out when the leakage occurs using standard techniques for leakage assessment  \cite{becker2013test, moradi2018leakage, bronchain2019multi}.
% Test Vector Leakage Assessment (TVLA) \cite{becker2013test}, $\chi^2$ test \cite{moradi2018leakage}, and Multi-tuple leakage detection \cite{bronchain2019multi}.  
Therefore it is challenging to pin-point the exact part of the micro-architecture causing this leakage.
\begin{figure}[t]
    \centering
    \includegraphics[width=.55\linewidth]{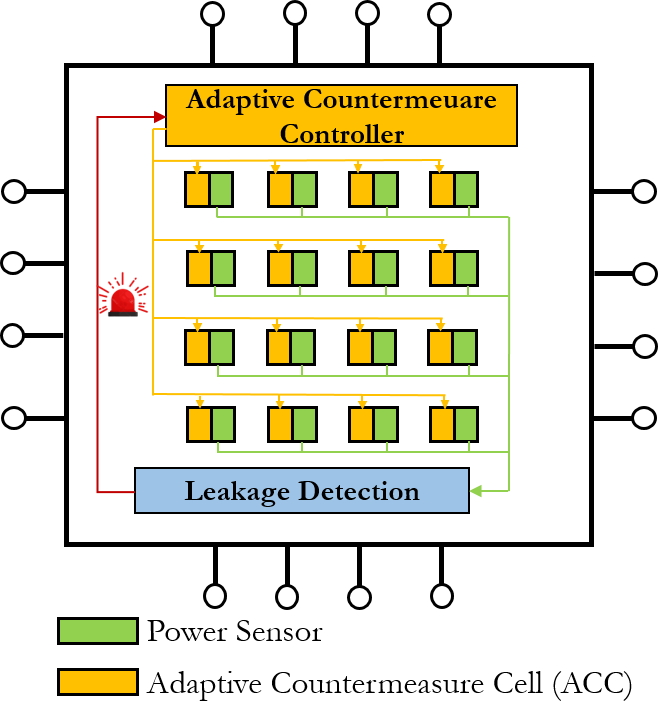}
    \caption{Proposed mechanism. Power sensors monitor the power. Leakage detection module predicts if a power SCL is happening based on the monitored power and alarms the adaptive controller. Adaptive countermeasure controller enables local countermeasures through ACCs.}
    \label{fig:method}
    \vspace{-.5cm}
\end{figure}
Current research efforts attempt to mitigate side-channel leakage from two ends. From the software side, researchers attempt to develop side-channel resistant software. This includes secure programming mechanisms \cite{barthe2017parallel, goudarzi2018secure}, and leakage-mitigating code-generation techniques \cite{wang2019mitigating}. From the hardware side, the hardware designers focus on building side-channel resistant processor designs and instruction set extensions to eliminate the dependencies of power consumption on the secret values and prevent accidental unmasking in masked software programs \cite{kiaei2020custom, kiaei2020domain}, or building up design-time (pre-silicon) side-channel evaluation \cite{yao2020architecture, yao2020verification}. 
% However, a considerable gap remains between these two fields. A software programmer is oblivious to the detailed design of hardware, and therefore cannot predict every single hardware effect of an instruction. At the same time, the hardware designer has no control over the software program running on the processor.
% Moreover, 
Software-only and hardware-only approaches are sub-optimal. The software designer will have to apply countermeasure to a substantial part of the code, if not the entire code, including the leaking section while being oblivious to the detailed design of the hardware. The processor hardware designer will have to build a completely new processor with corresponding countermeasures applied \cite{de2019protecting}. These approaches cause a large overhead in software execution time and code size, or in hardware area overhead and performance.

% Instead of mitigating the software or blindly adding hardware countermeasures, 
We propose to enable the processor system to monitor the side-channel leakage. We introduce a sensor that detects leakage in real-time and that triggers adaptive countermeasures. 
We present a proactive security mechanism based on hardware extensions.% within pipeline stages. 
The extensions provide a real-time monitor of the power consumption (using sensor cells in Fig.~\ref{fig:method}), detection of side-channel leakage (using leakage detection cells in Fig.~\ref{fig:method}), and real-time adaptive mitigation of detected side-channel leakage (using adaptive countermeasure cells (ACC) in Fig.~\ref{fig:method}). When the system is running, the proposed security mechanism (implemented as sensors) continuously monitors the leakage status of the system. Once a leakage is detected by the implemented sensors, the alarm signal will be set. This alarm signal has two functions: the alarm points out the precise leakage source in the hardware, and it triggers the adaptive countermeasure to mitigate the leakage.

\vspace{-.3cm}
\section{System Design}

We propose to build an insight into tracking, identifying, and mitigating side-channel leakage, while maintaining minimal implementation overhead. Using this methodology, the underlying hardware (processor) provides protection against power SCA for any program running on it. In our solution, then, the software programmer can be unaware of the power consumption of the hardware; the programmer may write code and generate the assembly using off-the-shelf compilers. The design time is therefore dramatically reduced, and the existing compiler optimizations are applied to provide a performance boost while ensuring resistance to power analysis attacks. 

The challenges facing the implementation of such a system are three-fold: detecting leakage, finding the source of leakage, and dynamically eliminating leakage. To address these challenges, our proposed real-time leakage detection system contains the following modules:  

\paragraph{In situ power measurement} Traditionally, power measurements were applied to the prototype of the chip. To achieve real-time leakage detection, we will implement an on-chip in situ power measurement based on power modeling. We also intend to study the possibility of utilizing existing on-chip sensors for the same purpose.  

% The challenge of this module is two-fold. First, to find a suitable model to simulate the power consumption of the leakage-critical parts of the hardware as close to the actual power measurement as possible.
% First, to accurately model the power consumption of the leakage critical parts of the hardware, i.e., finding the suitable models to simulate the power consumption as close to actual power measurement as possible. 
% Second, to achieve \textit{fine-grained} in terms of locality. Fine-grained in locality means we plan to model and monitor the power consumption of each critical component, e.g. every critical bit in a data-path register, separately. A granular model in the locality will lead us to the exact leakage point. 

%Second, to achieve \textit{fine-grained} in terms of both time and locality. \noteRed{Fine-grained in time means we want to achieve a high sampling rate for the modeled power simulation. Fine-grained in locality means we plan to model and monitor the power consumption of each critical component, e.g. every critical bit in a data-path register, separately.} Having a power model that is fine-grained in time helps us detect leakage regardless of any specific program. Furthermore, a granular model in the locality will lead us to the exact leakage point. 
%'high resolution'. Eg. why do you need a high sampling rate? Isn't a single number per instruction sufficient to decide of there is side-channel leakage or not?}

\begin{figure}
    \centering
    \includegraphics[width=.9\linewidth]{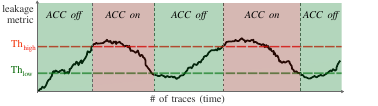}
    \caption{Each ACC turns on as soon as the leakage metric corresponding to its nearby power sensor passes the high threshold ($Th_{high}$). Once the leakage decrements to lower than $Th_{low}$, ACC will turn off.}
    \label{fig:adaptiveCountermeasure}
    \vspace{-.5cm}
\end{figure} 

\paragraph{Leakage detection}

After getting the real-time in situ power measurement, we will come up with the leakage evaluation mechanism and implement the corresponding hardware extension within the processor to achieve leakage detection. The key challenge in leakage detection is building a suitable leakage evaluation metric that fits well within the needs of real-time leakage detection and can be used for general-purpose programs. TVLA \cite{becker2013test} is currently the most popular leakage assessment technique, however, it requires categorization of inputs into fixed and random sets and therefore cannot be directly applied to our real-time leakage detection. Possible leakage metrics include power distribution-based characterization or machine learning-based metrics.
%Our leakage metric is based on characterizing the power consumption for random inputs off-line and comparing the power consumption in real-time with the characterized power consumption. By keeping the power consumption close to the power consumed by random inputs, we can conservatively prevent power SCL.

% The challenge of this part is to build a suitable leakage evaluation metric that fits well in the needs of real-time leakage detection. There exist statistical-based leakage evaluation methods such as TVLA which is widely used for off-line leakage detection. However, this method will be hard to fit into the requirements of real-time leakage detection and hardware implementation,  because it requires certain input patterns and mathematical calculations which are hard to implement in hardware. 

%MoreThe proposed hardware extension will achieve fine-grained leakage detection which will not only detect the leakage but also point out the leakage source once the leakage is detected. 

\paragraph{Adaptive countermeasure}

Our goal is to keep the leakage probability low; start the protection once the probability of leakage is higher than a threshold and stop it when the probability is lower than another (as shown in Fig.~\ref{fig:adaptiveCountermeasure}). This has the advantage that while the design is kept secure, only the necessary parts of the design are protected.

% The challenge of this module is adding and removing the precisely-targeted mitigation in hardware dynamically regarding the real-time leakage status of the system. This will cause hardware-triggered interrupts and stalls in order to apply protection to the leakage source and accordingly adjust the following parts of a program.
% There are possible two approaches to implement the adaptive countermeasures: hardware based and software based. In hardware based countermeasure, we can implement masking registers which can be enabled and disabled based on the real-time leakage status. In software-based countermeasure, we can enables hiding based protection such as insert random jitter into the system.  

% \paragraph{Prototype implementation}
% We will implement our proposed system on a four-stage implementation of RISC-V and prototype it on an FPGA. We will also study its effectiveness as an ASIC design and evaluate each one of the above-mentioned modules separately and as well as a whole system. Our lab is equipped with Riscure Inspector software package, a Teledyne oscilloscope with a sample rate as high as 40GS/s, Xilinx Zedboard, and Sakura-G FPGAs.
\vspace{-.3cm}
\section{Preliminary Implementation}

\begin{figure}
    \centering
    \includegraphics[width=.5\linewidth, angle =270]{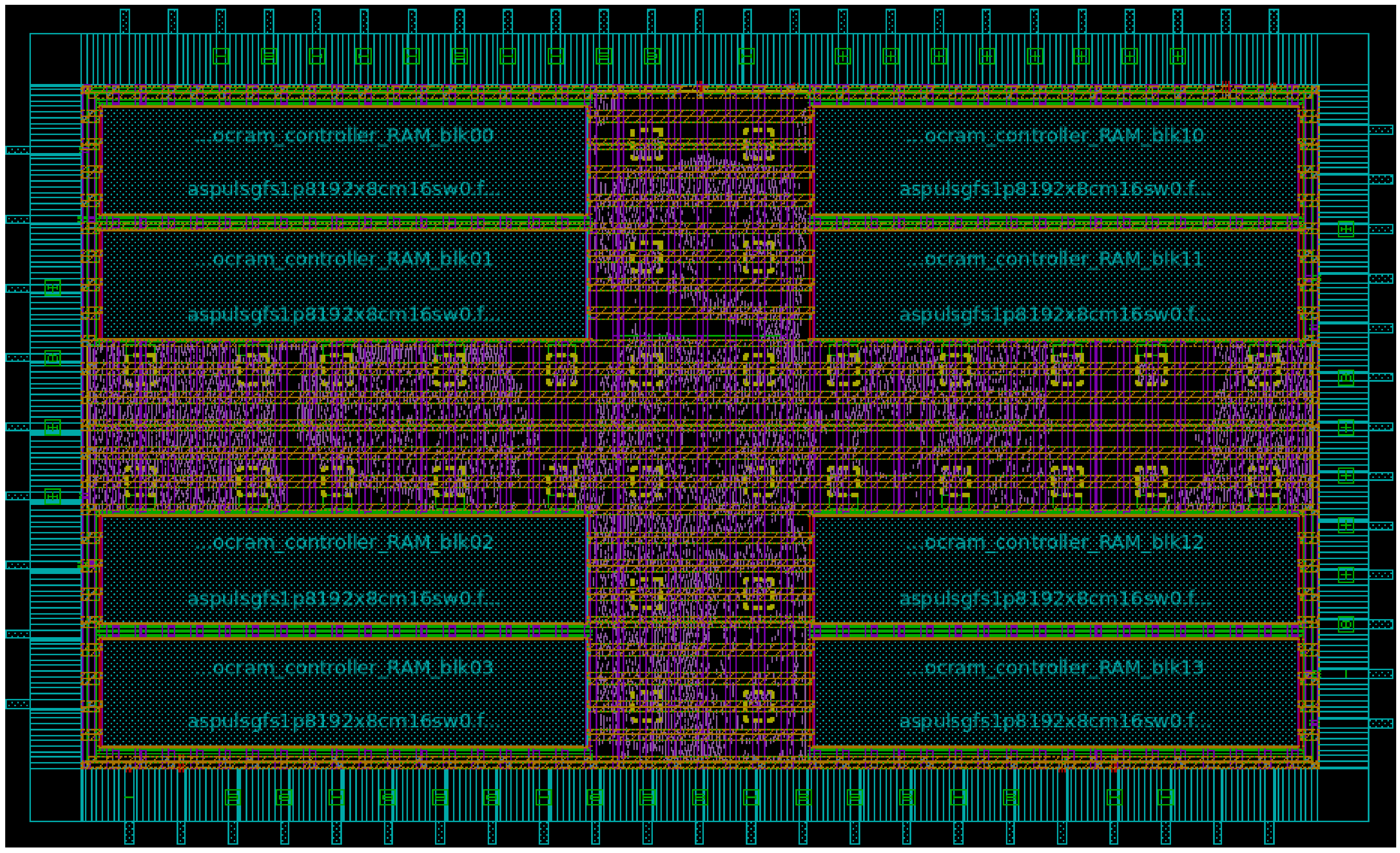}
    \caption{Layout of the chip, integrating the power sensors scattered throughout the layout. The yellow dashed squares highlight the power sensors.}
    \label{fig:picochip}
    \vspace{-.5cm}
\end{figure}

As the first step to build up the whole real-time system, we implemented the power sensors as hardware extensions to an SoC with a RISC-V processor and fabricated the chip with the CMOS 180nm technology for further testing. We chose digital Ring-Oscillators (RO) as our power sensors to be able to mirror the changes in the power consumption of the chip in through their oscillating frequency. The power sensors are added as co-processors in the SoC to communicate with the processor. In the chip layout, as shown in Fig.\ref{fig:picochip}, the power sensors are evenly distributed throughout the design to monitor the local power consumption in real-time while the chip is running a program.  

\vspace{-.3cm}
\section{Conclusion and Future Work}

We believe that our work is the first to build a mechanism that provides security to any algorithm and \textit{dynamically} applies corresponding countermeasures.
The presented system is expected to be very generic; once we have an implementation of such a system, we can securely run any algorithm on it. This is also the case for a protected processor implementation. Whereas, in a software protection realm, any new algorithm would need its own security implementation from the ground up, hence having a low genericness. All these advantages of our proposed method come at the cost of silicon area overhead which is still expected to be lower than that of a protected processor implementation.

\bibliographystyle{ieeepes}
% \bibliography{biblio}
{\footnotesize \bibliography{biblio}}

\end{document}